\documentclass[pdflatex,sn-basic,iicol]{sn-jnl}%

\usepackage{mathtools}
\usepackage{xspace}
\usepackage{hyperref}
\usepackage{tabulary}
\usepackage[inline,shortlabels]{enumitem} 
\usepackage[capitalise]{cleveref}

\jyear{2022}%

\newtheorem{theorem}{Theorem}%
\newtheorem{definition}[theorem]{Definition}

\newcommand{\calN}{\mathcal{N}}
\newcommand{\calM}{\mathcal{M}}
\newcommand{\lang}[1]{\mathcal{L}(#1)}
\newcommand{\eve}{Eve\xspace}
\newcommand{\adam}{Adam\xspace}
\newcommand{\out}[2]{\pi_{#1#2}}

\newcommand{\always}{\mathop{\mathtt{G}}}
\newcommand{\evtly}{\mathop{\mathtt{F}}}
\newcommand{\until}{\mathrel{\mathtt{U}}}
\renewcommand{\next}{\mathop{\mathtt{X}}}
\newcommand{\CC}{C\nolinebreak\hspace{-.05em}\raisebox{.4ex}{\tiny\bf +}\nolinebreak\hspace{-.10em}\raisebox{.4ex}{\tiny\bf +}}
\newcommand{\strix}{\texttt{Strix}}
\newcommand{\owl}{\texttt{Owl}}
\newcommand{\otus}{\texttt{Otus}}

\usepackage{soul}
\usepackage{xcolor}
\DeclareRobustCommand{\revone}[2]{#2}%
\DeclareRobustCommand{\revtwo}[2]{#2}%
\DeclareRobustCommand{\revthree}[2]{#2}%

\begin{document}

\title[SYNTCOMP '18--'21]{The Reactive Synthesis Competition (SYNTCOMP): 2018--2021}

\author*[1]{\fnm{Swen} \sur{Jacobs}}\email{jacobs@cispa.de}
\author*[2]{\fnm{Guillermo A.} \sur{P\'erez}}\email{guillermo.perez@uantwerp.be}

\author[3]{\fnm{Remco} \sur{Abraham}}
\author[4]{\fnm{V\'eronique} \sur{Bruy\`ere}}%
\author[5]{\fnm{Micha\"el} \sur{Cadilhac}}%
\author[6]{\fnm{Maximilien} \sur{Colange}}%
\author[4]{\fnm{Charly} \sur{Delfosse}}%
\author[3]{\fnm{Tom} \sur{van Dijk}}
\author[6]{\fnm{Alexandre} \sur{Duret-Lutz}}%
\author[7]{\fnm{Peter} \sur{Faymonville}}
\author[1]{\fnm{Bernd} \sur{Finkbeiner}}
\author[8]{\fnm{Ayrat} \sur{Khalimov}}%
\author[7]{\fnm{Felix} \sur{Klein}}
\author[9]{\fnm{Michael} \sur{Luttenberger}}
\author[9]{\fnm{Klara} \sur{Meyer}}
\author[6]{\fnm{Thibaud} \sur{Michaud}}%
\author[6]{\fnm{Adrien} \sur{Pommellet}}%
\author[6]{\fnm{Florian} \sur{Renkin}}%
\author[6]{\fnm{Philipp} \sur{Schlehuber-Caissier}}%
\author[10]{\fnm{Mouhammad} \sur{Sakr}}%
\author[9]{\fnm{Salomon} \sur{Sickert}}
\author[4]{\fnm{Ga\"etan} \sur{Staquet}}%
\author[4]{\fnm{Cl\'ement} \sur{Tamines}}%
\author[7]{\fnm{Leander} \sur{Tentrup}}
\author{\fnm{Adam} \sur{Walker}\textsuperscript{11}}%

\affil[1]{\orgname{CISPA Helmholtz Center for Information Security}, \orgaddress{\street{Stuhlsatzenhaus 5}, \city{Saarbr\"ucken}, \postcode{66123}, \country{Germany}}}
\affil[2]{\orgname{University of Antwerp -- Flanders Make}, \orgaddress{\street{Middelheimlaan 1}, \city{Antwerpen}, \postcode{2020},  \country{Belgium}}}
\affil[3]{\orgname{University of Twente}, \orgaddress{\country{The Netherlands}}}
\affil[4]{\orgname{University of Mons},
\orgaddress{
\country{Belgium}}}
\affil[5]{\orgname{DePaul University}, \orgaddress{
\country{USA}}}
\affil[6]{\orgname{Laboratoire de Recherche et D\'eveloppement de l'Epita}, \orgaddress{
\country{France}}}
\affil[7]{\orgname{Saarland University} \orgaddress{\country{Germany}}}
\affil[8]{\orgname{Universit\'e libre de Bruxelles}, \orgaddress{
\country{Belgium}}}
\affil[9]{\orgname{Technische Universit\"at M\"unchen}, \orgaddress{\country{Germany}}}
\affil[10]{\orgname{University of Luxembourg}, \orgaddress{
\country{Luxembourg}}}
\affil[11]{\orgname{Independent Researcher} \orgaddress{\country{Australia}}}

\abstract{We report on the last four editions of the reactive synthesis competition (SYNTCOMP 2018--2021). We briefly describe the evaluation scheme and the experimental setup of SYNTCOMP. Then, we introduce new
benchmark classes that have been added to the SYNTCOMP library and give an overview of the participants of SYNTCOMP. Finally, we present
and analyze the results of our experimental evaluations, including a ranking of tools with respect to
\revone{1}{quantity and quality --- that is, the total size in terms of logic and memory elements --- of solutions.}
}


\maketitle
\clearpage

\section{Introduction}
Reactive systems are systems that maintain a continuous interaction
with their environment. The act of automatically constructing such a system
from a given formal specification (or determining that no such system exists)
is called \emph{reactive synthesis}. The current definition of reactive
synthesis is usually attributed to Alonzo Church~\citep{church57,church64}.
In the last 60 years, several works have laid the theoretical foundations
that underpin all current synthesis algorithms for different instantiations of
reactive synthesis. Indeed, depending on the format in which a specification
for the reactive system is formalized, different synthesis problems arise. \revone{2}{For instance, the competition currently has three such specification formats: one for safety specifications defined by monitoring circuits, one for linear-temporal-logic
specifications, and one for parity-game specifications.}

Reactive synthesis has the potential to revolutionize
the way in which reactive systems are designed. This is due to the fact that a synthesized
system is correct by construction and therefore does not need to be tested nor
verified for correctness. Despite its potential, industry has not yet adopted it
nor the prototype tools implemented by academic researchers. This is in
contrast to other formal verification techniques such as model
checking~\citep{bk08,mc18}. With an aim at increasing the impact of reactive
synthesis in industry and improve the quality of synthesis tools, the Reactive
Synthesis competition (SYNTCOMP) was founded in 2014~\citep{syntcomp14}. In
short, the competition is designed to foster research into well-engineered,
scalable, and user-friendly synthesis tools. To realize this, the competition
organizers have proposed standards for benchmark formats, and maintain a
library of benchmarks with entries that remain challenging for
state-of-the-art tools. Most importantly, SYNTCOMP provides a dedicated and
independent platform for the objective comparison of synthesis tools.

SYNTCOMP has become an annual event associated with the International Conference
on Computer Aided Verification (CAV) and the Workshop on Synthesis (SYNT). The
organizational team of the competition has changed slightly from its inception: In
2019, Guillermo A. P\'erez joined the organization team and, since 2020, the
competition has an advisory committee that presently consists of Roderick
Bloem, Armin Biere, Salomon Sickert, Jean-Fran\c{c}ois Raskin, Bernd
Finkbeiner, and Ayrat Khalimov. Every year, the organizers publish a call for solvers and benchmarks on the website\footnote{\url{http://www.syntcomp.org/}} and via the associated mailing list\footnote{\url{https://lists.iaik.tugraz.at/cgi-bin/mailman/listinfo/syntcomp}}.

In this article we present the list of benchmark families that have been added
to the competition from 2018 to 2021 as well as the tools which participated
in the competition during the same years.
Finally, we also highlight the most interesting experimental results from
these editions of the competition and discuss the progress of synthesis tools observed in this time.

\section{Setup, Rules, and Execution}
We begin this section with a reminder on the foundations of the synthesis problem. For more details on reactive synthesis and (parity-)game solving, we refer the reader to the corresponding chapters in the books~\cite{mc18} and~\cite{ag11}.

\subsection{Synthesis and realizability}
To model the execution of a reactive system, we make use of infinite sequences:
An \emph{infinite word} $\alpha$ over an \emph{alphabet} $A$ is a function $\alpha\colon \mathbb{N}_{> 0} \to A$. Thus, we
write $\alpha(i)$ to refer to the $i$-th letter of $\alpha$. \revone{3}{We write $A^\omega$ to denote the set of all infinite words over $A$.}

\begin{definition}[Infinite-word automata]
  An ($\omega$-word) \emph{automaton} is a tuple $\calN =
  (Q,q_0,A,\Delta)$ where $Q$ is a finite set of states, $q_0 \in Q$ is the
  initial state, $A$ is a finite alphabet, and $\Delta \subseteq Q \times A
  \times Q$ is the transition relation. We assume that for all $p \in Q$ and all $a \in A$
  there exists $q \in Q$ such that $(p,a,q) \in Q$.
\end{definition}
The automaton
is said to be \emph{deterministic} if for all $p \in Q$ and all $a \in A$ we
have that $(p,a,q_1), (p,a,q_2) \in \Delta$ implies $q_1 = q_2$.
A \emph{run} of $\calN$ on a word $\alpha \in A^\omega$ is an infinite sequence $\rho =
q_0 \alpha(1) q_1 \alpha(2) \dots \in (Q\cdot A)^\omega$ such that
$(q_i,\alpha(i+1),q_{i+1}) \in \Delta$ for all $i \in \mathbb{N}$. 

Automata are paired with a condition that determines which runs are
\emph{accepting}.  For the competition, we
consider four such \emph{acceptance conditions}.
\begin{itemize}
  \item The \emph{safety} condition is defined with respect to a set $U
    \subseteq Q$ of \emph{unsafe states}. A run $\rho = q_0 a_1 q_1 a_2 \dots$
    is accepting for this condition, or \emph{safe}, if and only if for all $i \in
    \mathbb{N}$ we have that $q_i \not\in U$.
  \item The \emph{B\"uchi} condition is defined with respect to a set of
    \emph{B\"uchi} states $B \subseteq Q$. A run $\rho = q_0
    a_1 q_1 a_2 \dots$ is accepting for this condition if and only if for
    all $i \in \mathbb{N}$ there exists $j \ge i$ such that $q_j \in B$.
    That is, the run visits B\"uchi states infinitely often.
  \item The \emph{co-B\"uchi} condition is also defined with respect to a
    set $B \subseteq Q$ of states, which are in this case \emph{rejecting} states. 
    A run $\rho = q_0 a_1 q_1 a_2 \dots$ is
    accepting for the condition if and only if there exists $i \in \mathbb{N}$
    such that for all $j \ge i$ we have that $q_j \not\in B$.
    That is, rejecting states are visited only finitely often.
  \item The \emph{parity} condition is defined with respect to a
    \emph{priority function} $p\colon Q \to \mathbb{N}$. A run $\rho = q_0 a_1
    q_1 a_2 \dots$ is accepting for the parity condition if and only if the
    value $\liminf_{i \to \infty} p(q_i)$ is even. That is, the smallest priority that appears infinitely often along the run is even.
\end{itemize}
A word $\alpha$ is accepted by an automaton $\calN$ if it has a run on $\alpha$
that is accepting. In the sequel, it will sometimes be useful to consider
\emph{universal automata}. Such an automaton $\calN$ accepts a word $\alpha$
if all its runs on $\alpha$ are accepting.

We denote by $\lang{\calN}$ the \emph{language} of the automaton $\calN$, that
is, the set of words that $\calN$ accepts.

\subsubsection{Synthesis games and strategies}
\begin{definition}[Games]
  A \emph{(Gale-Stewart) game} on input and output alphabets $I$ and $O$,
  respectively, is a two-player perfect-information game played by \eve and
  \adam in rounds: \adam chooses an element $i_k \in I$ and \eve responds with
  an element $o_k \in O$. A \emph{play} in such a game is an infinite word
  $\langle i_1, o_1\rangle \langle i_2, o_2\rangle \dots \in (I\times O)^\omega$. 
\end{definition}
A game is paired with a \emph{payoff set} $P \subseteq (I \times
O)^\omega$ that determines who wins a play $\pi$. If $\pi \in P$ then
\emph{$\pi$ is winning for \eve}, otherwise it is \emph{winning for \adam}.

\begin{definition}[Strategies]
  A \emph{strategy} for \adam is a function $\tau\colon (I \times O)^*
  \to I$ which maps every (possibly empty) play prefix
  to a choice of input letter.  Similarly, a \emph{strategy for \eve}
  is a function $\sigma\colon (I \times O)^* I \to O$ which maps every play prefix
  and input letter to a choice of output letter.
\end{definition}
A play $\pi = \langle i_1, o_1\rangle \langle i_2, o_2\rangle \dots$ is
\emph{consistent with} a strategy $\tau$ for \adam if $i_k = \tau( \langle
i_1, o_1\rangle \dots \langle i_{k-1}, o_{k-1}\rangle)$ for all $k \in
\mathbb{N}$; it is consistent with a strategy $\sigma$ for \eve if $o_k =
\sigma( \langle i_1, o_1\rangle \dots \langle i_{k-1}, o_{k-1}\rangle,
i_{k})$.  A pair of strategies $\sigma$ and $\tau$ for \eve and \adam,
respectively, induces a unique play $\out{\sigma}{\tau}$ consistent with both
$\sigma$ and $\tau$.

In a game with payoff set $P$, the strategy $\sigma$ for \eve is a \emph{winning
strategy} if for all strategies $\tau$ for \adam it holds that
$\out{\sigma}{\tau} \in P$; the strategy $\tau$ for \adam is winning if for all
strategies $\sigma$ for \eve we have $\out{\sigma}{\tau} \not\in P$.

The realizability and synthesis problems are defined for games whose payoff sets
are given as the language of an automaton. We sometimes refer to these as
\emph{games played on automata}.
\begin{definition}[Realizability and Synthesis]
  Consider finite input and output alphabets $I$ and $O$, respectively, and
  an automaton $\calN$ with alphabet $I \times O$.
  The \emph{realizability problem} asks whether
  there exists a winning strategy for \eve in the game with payoff set 
  $\lang{\calN}$. The \emph{synthesis problem}
  further asks to compute and output such a strategy if one exists.
\end{definition}

\revtwo{1.a}{In SYNTCOMP, solvers are asked to solve the realizability and synthesis problems.
More precisely, the competition is organized into separate tracks according to different specification formats for games (see {\cref{sec:specifications}}), and in each track the participants are asked to solve these two problems.}

\subsubsection{Finite-memory strategies}
A \emph{finite-memory} strategy $\sigma$ for \eve in a game played on the
automaton $\calN = (Q,q_0,I \times O,\Delta)$ with finite input and output
alphabets $I$ and $O$ is a strategy that can be encoded as a
\emph{(deterministic) Mealy machine}, that is, a finite-state machine that
outputs a letter from \(O\) when given a letter from \(I\).  Formally, such a
machine is a tuple $\calM = (S,s_0,I, \lambda_u,\lambda_o)$ where $S$ is a
finite set of (memory) states, $s_0$ is the initial state,
$\lambda_u\colon S \times (Q \times I) \to S$ is the update function and
$\lambda_o\colon S \times (Q \times I) \to O$ is the output function. The
machine encodes $\sigma$ in the following way. For all play prefixes
$\langle i_1, o_1\rangle \dots \langle i_{k-1}, o_{k-1} \rangle$ and input
letters $i_k \in I$ we have that $\sigma(\langle i_1, o_1\rangle \dots \langle
i_{k-1}, o_{k-1} \rangle, i_k) = \lambda_o(s_k,i_k)$ where $s_{\ell + 1} =
\lambda_u(s_\ell,i_\ell)$ for all $\ell < k$. We then say the strategy
$\sigma$ has \emph{memory} $\lvert S \rvert$. In particular, when $\lvert S \rvert = 1$, we say the
strategy is \emph{memoryless} (the term \emph{positional} is also used in the
literature). For all games considered in the competition, it holds that there
exists a winning strategy for \eve in the game if and only if there exists a
memoryless winning strategy for \eve.

\subsection{Safety, parity, and linear-temporal specifications}
\label{sec:specifications}
Let $I$ and $O$ be finite input and output alphabets.  SYNTCOMP has tracks
corresponding to three different versions of the synthesis and realizability
problems. The \emph{safety tracks} correspond to these problems for games
played on deterministic automata with a safety acceptance condition;
the \emph{parity tracks}, to the same problems for games
played on deterministic automata with a parity acceptance condition. The
remaining tracks correspond to games whose payoff set is given in the form of
a \emph{linear-temporal-logic} (LTL) formula. We explain the connection between LTL formulas and infinite-word automata in the following.

\subsubsection{LTL-defined payoff sets}
LTL~\citep{pnueli77} is a logic that allows one to naturally specify time
dependence among events that make up the formal specification of a system.
Formulas in LTL are constructed from a set $P$ of atomic propositions, the
usual Boolean connectives, and temporal operators $\next, \evtly, \always,
\until$ which intuitively correspond to ``next'', ``eventually'', ``always'',
and ``until'' in English. Formally, LTL formulas conform to the following
syntax:
\[
  \varphi \Coloneqq a \in P \mid \varphi \land \varphi \mid
  \lnot \varphi \mid \next \varphi \mid \evtly \varphi \mid
  \always \varphi \mid \varphi \until \varphi
\]
with derived operators such as implication defined as usual.
For instance, the formula $\always(\mathit{req} \rightarrow \evtly
\mathit{grant})$, over atomic propositions $\mathit{req}$ and
$\mathit{grant}$, can be read as ``it is always the case that if there is a
request then eventually it is granted''. We refer the reader to the book
by~\citet{bk08} for the formal semantics of LTL. In the context of words over an
input-output alphabet $I \times O$, the atomic propositions can be assumed to
be an encoding of letters in the alphabet. That is, the truth value of the
propositions is defined for each letter.

It is well known that the set $\mathrm{Words}(\varphi)$ of all words
satisfying a given LTL formula $\varphi$ can be ``compiled'' into an
infinite-word automaton. For instance, one can construct (in exponential time)
a non-deterministic automaton $\calN$ with a B\"uchi acceptance condition such that
$\lang{\calN} = \mathrm{Words}(\varphi)$~\citep{vw84}. One can also construct
(in doubly-exponential time) a deterministic automaton $\calN$ with a parity
acceptance condition with the same property~\citep{safra88,piterman07}. 

The
\emph{LTL tracks} of the competition correspond to synthesis and
realizability problems that can for example be solved by playing games 
on a deterministic parity automaton compiled from a given LTL formula, or 
by other solutions that rely on different automata constructions.
In particular, algorithms
for both problems exist which avoid the expensive construction of a deterministic automaton and can work with non-deterministic automata. 

\subsubsection{Specification formats}
We now briefly touch on the encoding used by the competition to represent the
input for the safety, parity, and LTL tracks.
\begin{description}
    \item[Safety specifications.] To represent (the transition relation of a) deterministic
    automata with a safety acceptance condition, we use And-Inverter
  Graphs (AIGs). In turn, to encode AIGs, we use an extended version of the AIGER format~\citep{aiger}. The latter is the standard format in the hardware model checking competition~\citep{hwmcc}. The main reason that the basic format has to be extended is to allow for the partitioning of the alphabet into $I$ and $O$~\citep{DBLP:journals/corr/Jacobs14}.
    \item[Parity specifications.] For automata with a parity acceptance condition, we use an extended version of the Hanoi Omega-Automata (HOA) format~\citep{DBLP:conf/cav/BabiakBDKKM0S15}. The HOA format is a flexible exchange format for infinite-word automata. Just like the AIGER format, extending it is necessary to be able to include the partitioning of alphabet into $I$ and $O$~\citep{DBLP:journals/corr/abs-1912-05793}.
    \item[LTL specifications.] Finally, to represent LTL specifications we use the Temporal Logic Synthesis Format (TLSF)~\citep{tlsf}. \revtwo{1.d}{TLSF allows to define \emph{families} of LTL specifications via parameters.} Additionally, it allows to use high level constructs, such as sets and functions, to provide a compact and human-readable representation. 
\end{description}

\subsubsection{Output format}
For the synthesis tracks, tools are expected to produce a strategy if the specification is realizable. AIGER is the format used by the competition to encode the Mealy machine implementing the strategy. In this case the standard AIGER format is sufficient.

\subsection{Rules}

All tracks are divided into subtracks for
realizability checking and synthesis, and into two execution modes: sequential (using a single core of
the CPU) and parallel (using up to 4 cores).
Every tool can run in up to three configurations per subtrack and execution mode. Before the competition, all tools are tested on a small benchmark set, and authors can submit bugfixes if problems are
found. Tools submitted by the organizers are not allowed to submit bugfixes.

\paragraph{Disqualification}
\revtwo{3}{During the competition, \textbf{no erroneous results are allowed}. For the realizability subtracks, this is easy to check. For the synthesis subtracks, we model check all synthesized strategies. In the exceptional case that the output of a tool cannot be model checked, e.g. because of it being too large for it to be analyzed within reasonable time and space, then it is assumed to be correct.\footnote{In practice, this only happened a handful of times.}}

\paragraph{Extraordinary comparisons}
\revtwo{5}{There are two cases in which an unofficial comparison run is launched. First, tools or bugfixes may be submitted after the competition is over. Second, for the purpose of including in the comparison tools that are no longer maintained by their authors, the organizers of SYNTCOMP sometimes modify or compile a (previous version) of the tool themselves. When reporting the results of a competition, we include tools that fit these two cases but refer to their results as being \textit{hors concours}. That is, they did not official participate in the competition.}

\subsubsection{Ranking schemes} In all tracks, there is a ranking based on the number of correctly solved problems within a $3600$s timeout per benchmark. 
In the synthesis tracks, correctness of the solution additionally has to be
confirmed by a model checker.
Moreover, in synthesis tracks there is a ranking based on the quality of the solution, measured by
the number of gates in the produced AIGER circuit. To this end, the size of the solution is compared to
the size $r$ of a reference solution. A circuit of the same size is rewarded $2$ points, and smaller or
larger solutions are awarded more or less points, respectively (see, e.g.~\citet{syntcomp17}, for more details).

\subsubsection{Selection of benchmarks} In 2018, benchmarks were selected according to the same scheme as in previous
years, based on a categorization into different classes (again, see~\citet{syntcomp17}). 
\revtwo{1.b}{From 2019 onward, all available benchmarks in the SYNTCOMP library (see {\cref{sec:benchmarks}}) were used. In the LTL track, some of the TLSF-encoded benchmarks represent a family of benchmarks that can be scaled up in one or more parameters. For these, more instances are generated whenever all the existing ones are considered ``too easy'' for more than one tool.}

\subsection{Execution}
In 2018, SYNTCOMP was run at Saarland University.  Benchmarking was again organized on the EDACC platform~\citep{DBLP:conf/lion/BalintDGGKR11}.
Since 2019, the competition has been run on StarExec~\citep{DBLP:conf/cade/StumpST14}. This has had a few consequences. The main such consequence is that not all legacy tools were successfully migrated: some of them relied on deprecated packages while others were implemented in languages with limited compiler support. Furthermore, such tools lacked an active maintainer. 

Our use of StarExec has significantly simplified the organizational effort, while admittedly raising the entry threshold for participants a bit, since they can, and must, themselves make sure that their code runs on the competition servers. \revtwo{D}{In practice, all tools that were still actively maintained (more precisely, tools for which a Ph.D. student worked on the tool) were updated and migrated to StarExec. Any reduction in the number of participating tools from 2018 to 2019 is instead due to a number of projects ending and Ph.D. students graduating around the time.} For an up-to-date overview of the specifications of the StarExec service, we refer the reader to its website\footnote{\url{https://www.starexec.org/}} and wiki.\footnote{\url{https://wiki.uiowa.edu/display/stardev/User+Guide}}

\section{Benchmarks}
\label{sec:benchmarks}

\revtwo{1.c}{Since the inception of SYNTCOMP, benchmarks in the different specification formats have been collected in the SYNTCOMP library\footnote{\url{https://github.com/SYNTCOMP/benchmarks}}.
The benchmarks in the SYNTCOMP library mainly come from three sources:}
\begin{enumerate*}[i)]
\item participants are invited to submit benchmarks from their own research,
\item benchmarks from the literature are translated into our standard format, and
\item we use different techniques to automatically generate additional benchmarks.
\end{enumerate*}

In the following, we shortly comment on families of benchmarks that were added to each one of the tracks during the relevant years.

\subsection{LTL}
Between 2018 and 2019, several \emph{temporal-stream logic} (TSL) benchmarks were submitted to the LTL tracks of SYNTCOMP. TSL is a new temporal logic that allows to separate control and data~\citep{DBLP:conf/cav/Finkbeiner0PS19}. Among others, the logic can be used to specify components of games implemented for FPGAs~\citep{DBLP:conf/fmcad/GeierH0F19} and to specify functional reactive programs~\citep{DBLP:conf/haskell/Finkbeiner0PS19}. Importantly, bounded TSL specifications can be translated into LTL specifications --- this is how the benchmarks are obtained.

In addition to the TSL benchmarks, Felix Klein also submitted LTL benchmarks based on hardware-component specifications and an encoding of an ``infinite duration tic-tac-toe". In 2020 and 2021, TLSF files encoding families of LTL specifications were used to generate more challenging benchmarks, i.e., larger parameter values were used to generate more difficult instances of each family to gauge how well tools scale w.r.t. the parameters.

In \autoref{tab:ltlbenchs} we summarize some statistics of interest regarding the benchmarks used for all relevant editions of the LTL tracks. All data used to generate the table was fetched from \url{https://syntcomp.react.uni-saarland.de/} and \url{https://github.com/SYNTCOMP/benchmarks/releases}.

\begin{table*}
\footnotesize
\centering
\caption{Statistics regarding the LTL benchmarks: all generated formulas are
fully parenthesized. Thus, by formula length we mean the number of characters;
by formula depth, the maximal
nesting of temporal operators (obtained using Spot's \texttt{ltlfilt} utility). Finally, all values are presented in the
format ``$\mathrm{min} \leq \mathrm{median,mean} \leq
\mathrm{max}$''.\label{tab:ltlbenchs}}
\begin{tabulary}{\textwidth}{Lllllll}
  Year & No. of bench. & Length of LTL formula & Formula depth & No. of inputs & No. of outputs & Solved by winner\\
  \midrule
  2018 & $286$&$28\leq349, 682\leq9358$&$1\leq2, 3\leq19$&$1\leq4, 5\leq25$&$1\leq2, 3\leq14$ & $93$\%\\
  2019 & $346$&$28\leq464, 1031\leq12843$&$1\leq2, 3\leq37$&$1\leq4, 5\leq25$&$1\leq2, 4\leq37$ & $94$\%\\
  2020 &
  $346$&$28\leq464, 1031\leq12843$&$1\leq2, 3\leq37$&$1\leq4, 5\leq25$&$1\leq2, 4\leq37$ & $98$\%\\
  2021 & $942$&$23\leq896, 2031\leq330512$&$0\leq3, 4\leq37$&$1\leq4, 7\leq70$&$1\leq3, 5\leq64$ & $90$\%\\
\end{tabulary}
\end{table*}

\subsection{Safety}
In 2019, random benchmarks in the extended AIGER format were systematically generated by Mouhammad Sakr by uniformly sampling
from the set of all specifications with given values for the number of
(un)controllable inputs and latches. The techniques used for this were
subsequently improved and described by~\citet{DBLP:conf/cav/JacobsS20}.

\subsection{Parity games}
In 2020, Spot~\citep{duret.16.atva2} was used to generate deterministic parity automata from some LTL
benchmarks. It is important to mention that, perhaps because this translation is doubly
exponential in general, not many specifications could be translated.
Furthermore, the ones that did yield a deterministic parity automaton resulted
in small automata. To be precise $121$ benchmarks were used for that first edition of the parity-game track and they were all generated as previously mentioned.

In 2021, more benchmarks were translated as described above (for a total of $303$ benchmarks generated in that way). Also, ($217$) combinatorially hard
benchmarks were generated using the PGSolver parity-game suite~\citep{DBLP:conf/atva/FriedmannL09} and then
translated to the extended HOA format.

\section{Updated participating tools}\label{sec:updated-tools}
In this section we give an overview of the participants of the 2018--2021 editions of SYNTCOMP. We mostly focus on the participants of the parity and LTL tracks. \revone{4}{The safety track had minimal active participation: in 2018, the only tool that received an update was Simple BDD Solver, and the only new tool was LazySynt, which participated \textit{hors concours}. No updates or new tools were submitted to the safety track after 2018.}

\begin{table*}
\footnotesize
\centering
\caption{\revthree{5}{Participation years, authors, and links to source code for tools mentioned in this article}}
\begin{tabulary}{\textwidth}{Llll}
  Tool \& participation years & Developers & URL & Section\\
  \midrule
  Acacia bonsai & Cadilhac et al. & \url{https://github.com/gaperez64/acacia-bonsai} & \autoref{sec:acaciabonsai} \\
  BoSy ('17, '18) & Faymonville et al. & \url{https://www.react.uni-saarland.de/tools/bosy/} & \autoref{sec:bosy} \\
  BoWSer ('17, '18) & Finkbeiner et al. & \url{https://www.react.uni-saarland.de/tools/bowser/} & \autoref{sec:bowser} \\
  Knor ('20--'21) & Tom van Dijk & \url{https://github.com/trolando/knor} & \autoref{sec:knor} \\
  LazySynt & Adam Walker & \url{https://github.com/mhdsakr/Lazy-Safety-Synthesis} & \autoref{sec:lazysynt} \\
  Ltlsynt ('17--'21) & Renkin et al. & \url{https://spot.lrde.epita.fr/ltlsynt.html} & \autoref{sec:ltlsynt} \\
  \otus{} (2021) & Abraham et al. & \url{https://doi.org/10.5281/zenodo.5046346} & \autoref{sec:otus} \\
  Party/Kid \& \texttt{sdf} ('17, '18, '21)  & Ayrat Khalimov & \url{https://github.com/5nizza/sdf-hoa} & \autoref{sec:sdf} \\
  Simple BDD Solver ('14--'21) & Sakr et al. & \url{https://github.com/adamwalker/syntcomp} & \autoref{sec:simplebddsolver} \\
  SPORE (2021) & Bruy\`ere et al. & \url{https://github.com/Skar0/spore} & \autoref{sec:spore} \\
  \strix{} ('18--'21) & Meyer et al. & \url{https://github.com/meyerphi/strix} & \autoref{sec:strix} \\
\end{tabulary}
\end{table*}

Some of the tool descriptions below may refer to specialized techniques. We refer the interested reader to the publications cited in the text for further details on such techniques. After the overview of the participants we conclude this section with a classification of the LTL-synthesis tools based on techniques, data structures, and algorithms.

\subsection{Acacia bonsai}\label{sec:acaciabonsai}
Acacia bonsai, the spiritual successor of Acacia+~\citep{bbfjr12},
participated hors concours in the 2021 edition of the LTL-realizability track.
It implements \emph{downset}-based algorithms (i.e. algorithms that manipulate downward closed sets) that avoid constructing a
deterministic automaton for the given LTL specification. Instead, the downsets
are used to efficiently store sets of states in an on-the-fly determinization process. These algorithms were
  introduced by Filiot et al.~in the 2010s and implemented in the tools Acacia
  and Acacia+ in C and Python~\citep{bohy14}.  Acacia bonsai is a complete rewrite of Acacia in \CC20, articulated
  around \emph{genericity} (that is, a library of downset functions with generic type parameters) and leveraging modern techniques for better performance.
  These techniques include compile-time specialization of the algorithms, the 
  use of SIMD registers to store vectors, and several preprocessing steps, some
  relying on efficient Binary Decision Diagram (BDD) libraries.  It also includes
  different data structures to store downsets such as \emph{k-d trees}, a useful data structure for organizing points in a k-dimensional space (see, e.g. \citet{bcko08}).
  
  It is worth mentioning that, to compile the
input LTL formula into an automaton, Acacia bonsai uses Spot~\citep{duret.16.atva2}.

\subsection{BoSy}\label{sec:bosy}
BoSy was updated by P. Faymonville, B. Finkbeiner and L. Tentrup in 2018 and competed in both the realizability and the synthesis track. To detect realizability, BoSy translates the (complement of the) LTL specification into a safety automaton by bounding the number of visits to B\"uchi
states. The resulting safety game is solved by SafetySynth. For synthesis, BoSy relies on an encoding into quantified Boolean formulas (QBF). A full account of the algorithms implemented in the tool is given by~\citet{DBLP:conf/cav/FaymonvilleFT17}.
Two configurations of BoSy competed in SYNTCOMP 2018: configuration (basic) and configuration
(opt), where the latter further improves the size of the strategy by encoding the existence of an AIGER
circuit representing the strategy directly into a QBF query. Both configurations support a parallel mode,
if more than one core is available.

\subsection{BoWSer}\label{sec:bowser}
BoWSer was updated by B. Finkbeiner and F. Klein in 2018. It implements different extensions of the \emph{bounded
synthesis} approach that solves the LTL synthesis problem by first translating the complement of the specification into a B\"uchi automaton, and then encoding acceptance of a transition system with bounded
number of states into a constraint system, in this case a propositional satisfiability (SAT) problem. The details of all encodings are described in \citet{DBLP:conf/cav/FinkbeinerK16}.

Compared to 2017, a number of small improvements to speed up computations were implemented, and an experimental preprocessor to simplify LTL formulas has been added.
The sequential configurations of the tool spawn multiple threads that are executed on a single CPU
core. The parallel configurations are mostly the same as the sequential ones, but use a slightly different
strategy for exploring the search space of solutions.

\subsection{Knor}\label{sec:knor}
Knor is a BDD-based solver for parity
specifications first submitted by T. van Dijk in 2020. It leverages the Sylvan BDD
package~\citep{DBLP:journals/sttt/DijkP17} and the Oink parity-game
solver~\citep{vandijk.18.tacas}.

Knor implements a translation from HOA
to a symbolic parity automaton encoded using BDDs. Importantly, the chosen variable ordering encodes first source states, then uncontrollable inputs, controllable inputs, and finally the target states. The resulting symbolic parity automaton can then be treated in two ways: First, it can be solved directly using a symbolic parity game algorithm~\citep{DBLP:journals/corr/abs-2009-10876} optimized to utilize the aforementioned variable ordering. This solution immediately yields a controller that can be dumped as an AIG. Alternatively, the symbolic parity automaton can be output as an explicit parity game. (Note that the previous encoding into BDDs might have reduced the size of the automaton, exponentially in some cases.) Such an explicit parity game can then
solved by any of the algorithms implemented in Oink. This last option does not yet support synthesis, only realizability.

For the first solution, synthesis is realized using a na\"ive construction
of an AIG realizing the Mealy controller, based on a simple application of the Shannon expansion of the BDD-encoded functions. 

\subsection{LazySynt}\label{sec:lazysynt}
The \emph{Symbolic Lazy Synthesis} (LazySynt) tool
was submitted in 2018 by M. Sakr and S. Jacobs. It participated \textit{hors concours} in the safety-synthesis
track. In contrast to the classical BDD-based algorithm and the SAT-based methods implemented in
Demiurge~\citep{DBLP:conf/vmcai/BloemKS14,DBLP:conf/date/SeidlK14}, LazySynt implements a combined forward-backward search that is embedded into a refinement loop, generating candidate solutions that are checked and refined with a combination of backward
model checking and forward generation of additional constraints~\citep{DBLP:conf/cav/JacobsS20}.

\subsection{Ltlsynt}\label{sec:ltlsynt}
\begin{table*}
\footnotesize
\centering
\caption{Versions of Spot on which \texttt{ltlsynt} submissions to SYNTCOMP
were based.\label{tab:summary}}
\begin{tabulary}{\textwidth}{llJ}
  Year & Version  & Main changes in \texttt{ltlsynt} \\
  \midrule
  2017 & pre-2.4 & first implementation \\
  2018 & 2.5.3 & optimizations to determinization, and game solving; incremental determinization
  approach \\
  2019 & 2.7.4 &  (bugged) latest appearance record (LAR); improved LTL translation; incremental determinization removed \\
  2020 & 2.9 & reimplemented LAR, split, and game solving; parity minimization \\
  2021 & 2.9.7 & input decomposition; strategy simplification; specialized strategy construction for some LTL fragments\\
\end{tabulary}
\end{table*}

The program \texttt{ltlsynt}, introduced to SYNTCOMP in 2017~(see \autoref{tab:summary}), is part of Spot~\citep{duret.16.atva2}.  It
relies on a translation of the LTL specification to a parity game
whose winning strategy is then encoded as an AIGER circuit.  The
version submitted to the 2021 edition features the following
improvements~\citep{renkin.21.synt}:
\begin{itemize}
  \item  A decomposition of the input specification when possible~\citep{finkbeiner.21.nfm}.
  \item An LTL translation to Deterministic Emerson-Lei Automata (DELA) that
handles various simplifications: splitting the input formula in a manner
similar to the \texttt{delag} tool~\citep{muller.17.gandalf},
detecting \emph{obligation} subformulas \citep{esparza.18.lics},
relying on weak automata and suspendable properties~\citep{babiak.13.spin}.
  \item An SCC-based paritization algorithm~\citep{renkin.20.atva} for DELA
that relies on a \emph{color appearance record}.
  \item A transition-based parity game solver adapted
from~\citet{vandijk.18.tacas}, supporting (non-recursive) SCC decomposition
and parity compression.
  \item Optimization of the winning strategy through a variant of Spot's
simulation-based reduction based on BDD signatures
or an improvement of a SAT-based minimization algorithm for
Incompletely Specified Mealy Machines~\citep{renkin.22.forte}.
\end{itemize}

\subsection{Otus}\label{sec:otus}

\otus~\citep{essay87386} is a tool for LTL synthesis using
symbolically-represented parity automata and games. It proceeds as follows:
the LTL formula is decomposed into a Boolean combination of \emph{simpler}
formulas, these formulas are separately translated to BDD-encoded
deterministic automata, and then recomposed by computing the (deterministic)
union and intersection on the BDD-representation. Concretely, \otus{}
\begin{enumerate}
    \item makes
use of the $\Delta_2$-normalisation and the translation to deterministic
(co-)B\"uchi automata found in \citep{DBLP:conf/lics/SickertE20} implemented
by \owl~\citep{DBLP:conf/atva/KretinskyMS18},
    \item computes the symbolic
representation of a deterministic Rabin automaton by union and intersection,
and
\item applies a symbolic implementation \citep{paritizing} to obtain a
parity automaton. 
\item This symbolic automaton is reinterpreted as a parity game that is then 
solved by a symbolic implementation of the distraction fix-point iteration
\citep{DBLP:journals/corr/abs-1909-07659,DBLP:journals/corr/abs-2009-10876}.
\end{enumerate}

In order to speed up
the BDD-operations, \otus{} makes use of the Sylvan BDD
package~\citep{DBLP:journals/sttt/DijkP17}.

\subsection{Party/Kid and {\tt sdf}}\label{sec:sdf}
Party/Kid and {\tt sdf}, submitted by A. Khalimov, implement variants of symbolic bounded synthesis \citep{E11unbeast,E10}.
Both tools run two tasks, realizability and unrealizability, in parallel. Below, we describe how the realizability check is done. Unrealizability can be checked in a similar fashion by adequately modifying the given LTL formula.

First, the tool translates the given LTL
formula into a universal co-B\"uchi automaton (UCW) using the TLSF-manipulation tool {\tt syfco}~\citep{tlsf} and the Spot automata library~\citep{duret.16.atva2}.
Then, it iterates over increasing bounds
on the number of visits to final states of the UCW:
given such a bound, it translates the UCW into a universal safety automaton.
The universal safety automaton is then encoded into a safety game in a BDD-based representation (using the CUDD library~\citep{cudd}),
where each state of the universal automaton gets a separate variable in the BDDs,
thus avoiding explicit determinization.
The game is then solved using the standard fix-point algorithm.
The strategy extraction is also standard and does not use any third-party tools.

Over the years 2017, 2018, 2021, three versions participated, with only technical differences.
The latest version ({\tt sdf}, 2021) is written in \CC.

\subsection{Simple BDD Solver}\label{sec:simplebddsolver}
An update of Simple BDD Solver, submitted by Adam Walker, competed in the 2018 safety-realizability track.
Simple BDD Solver implements the classical BDD-based fixpoint algorithm for safety games~\citep{syntcomp14}. In sequential mode, it runs in three configurations, two of which are based on an abstraction-refinement approach
inspired by~\citet{DBLP:journals/iandc/AlfaroR10}, and one without any abstraction. All
three implement many important optimizations. These configurations are the same as in 2017.
Additionally, three new configurations were entered for the parallel mode. These run different
portfolios of the algorithms in the sequential mode.

\subsection{SPORE}\label{sec:spore}
SPORE is a prototype tool designed to assess the viability of using
\emph{generalized parity games} for LTL realizability These are games played on automata with multiple priority functions requiring all of the corresponding parity conditions to be satisfied. The input LTL formula is
first decomposed into a conjunction of sub-formulas, which are in turn
translated into deterministic parity automata, composed into a synchronised
(generalized parity) product automaton, and finally translated into a
generalized parity game using the {\tt tlsf2gpg}\footnote{See
\url{https://github.com/gaperez64/tlsf2gpg}} tool.  The game is then solved
using a combination of the recursive algorithm for generalized parity games
\citep{ChatterjeeHP07} and incomplete polynomial-time algorithms, called
partial solvers, presented in \citet{BruyerePRT19}.  SPORE contains an
explicit and a symbolic implementation of those algorithms, the latter relying
on BDDs to represent the game arena.  The version
presented in 2021 is implemented in Python with dd\footnote{See
\url{https://github.com/tulip-control/dd}} as a library to manipulate BDDs.

\subsection{Strix}\label{sec:strix}

\strix{} is a tool for LTL synthesis using transition-based deterministic parity automata (tDPW) and parity games as intermediate steps. Since its inception \citep{DBLP:conf/cav/MeyerSL18}, it proceeds in four stages: 1) Formula Rewriting and Decomposition, 2) Automaton Construction, 3) Winning Strategy Computation, and 4) Controller Extraction. Note that stages 2 and 3 run in parallel, are on-the-fly, and exchange information, such that the automaton construction can be focused on critical states and early termination is possible. 

The improvements applied for SYNTCOMP 2019 and 2020 are described by \cite{DBLP:journals/acta/LuttenbergerMS20} and include a refined controller extraction stage and updates to the LTL-translations based upon the work of \cite{DBLP:journals/jacm/EsparzaKS20} provided by \owl~\citep{DBLP:conf/atva/KretinskyMS18}. For SYNTCOMP 2021 the following major changes have been applied \citep{meyer.21.synt}:

\begin{itemize}
	\item LTL formulas are now translated to transition-based deterministic Emerson-Lei automata (tDELA) by combining constructions ($\Delta_2$-normalisation, direct translation to deterministic automata) from  \cite{DBLP:conf/lics/SickertE20} with a product construction adapted from \cite{muller.17.gandalf}. Then either a tDELA-to-tDPW construction based on Zielonka-trees or the Alternating Cycle Decomposition is applied (\cite{DBLP:conf/icalp/CasaresCF21,DBLP:conf/tacas/CasaresDMRS22}).
	\item The strategy iteration algorithm, by \cite{DBLP:journals/corr/abs-0806-2923}, for solving parity games, has been replaced by a Rust implementation of the distraction fixpoint iteration algorithm (DFI) \citep{DBLP:journals/corr/abs-1909-07659}.
\end{itemize}

\subsection{Classification of LTL tools}\label{sec:class-ltl}
To summarize some of the tool descriptions given above, the participants of the competition can be classified as implementing one of two approaches.
\begin{description}
    \item[Bounded synthesis.] That is, they translate the LTL specification
      into a universal co-B\"uchi automaton. Then, for an increasingly larger
      bound $b \in \mathbb{N}$, either (i) the automaton is turned into a
      universal safety automaton by allowing at most $b$ visits to
      rejecting states, and then determinized, or (ii) the search is
      limited to winning strategies encoded as a Mealy machine with at most
      $b$ memory states. The resulting problems are then either solved by
      standard techniques for safety games, or by encoding them into a
      constraint system and employing a SAT-, QBF- or SMT-solver.
    \item[The Owl-Pig approach.] We coin the class name \emph{Owl-Pig} to
      refer to solvers that first translate the LTL specification into a
      deterministic parity automaton using one of several algorithms described
      in the works of S. Sickert et al. and implemented in the \owl{} tool
      (see \autoref{sec:strix}). Then, they solve the resulting parity game
      using one of several algorithms described in the works of T. van Dijk et
      al. and implemented in the tools Oink and Knor (the sound a pig makes in
      English and Dutch respectively).
\end{description}
Tools that implement the bounded synthesis approach mostly differ in choice of programming language, (symbolic) data structure or logic encoding, and techniques for \emph{counter reduction}, that is, the detection of rejecting states for
  which knowing whether they have been visited or not is sufficient (i.e.
  instead of storing the exact number of visits).

For the Owl-Pig approach, generating in full the deterministic parity automaton is clearly the bottleneck. Hence, optimizations, types of automata used as intermediate steps, symbolic encodings, and on-the-fly constructions are the main differences among those tools. We present our classification and summarize some noteworthy differences between them in \autoref{tab:bsynthesis} and \autoref{tab:owlpig}. We concede that our classification is not perfect: for instance, SPORE does not really implement the Owl-Pig approach, yet we classify it as such since it does go through a similar $2$-stage algorithm: translate the specification to a deterministic automaton and solve a game.

\begin{table}
\footnotesize
\centering
\caption{Differences between bounded synthesis tools.\label{tab:bsynthesis}}
\begin{tabulary}{\textwidth}{Ll}
  Tool & Symbolic data structures/encodings\\
  \midrule
  Acacia bonsai & Antichains \\
  BoSy & BDDs, SAT, QBF \\
  BoWSer & SAT  \\
  Party/Kid \& \texttt{sdf} & BDDs \\
\end{tabulary}
\end{table}

\begin{table*}
\footnotesize
\centering
\caption{Differences between Owl-Pig tools.\label{tab:owlpig}}
\begin{tabulary}{\textwidth}{Lll}
  Tool & Automata used & Game-solving algorithms \\
  \midrule
  Ltlsynt & Det. Emerson Lei, Parity & Transition-based parity-game solver~\citep{vandijk.18.tacas}\\
  \otus{} & BDD-encoded B\"uchi, Rabin, Parity & Symbolic DFI~\citep{DBLP:journals/corr/abs-1909-07659,DBLP:journals/corr/abs-2009-10876}\\
  SPORE & Generalized Parity & Recursive + partial
  solvers~\citep{ChatterjeeHP07,BruyerePRT19}\\
  \strix{} & Det. Emerson Lei, Parity & Distraction fixpoint (DFI)~\citep{DBLP:journals/corr/abs-1909-07659}\\
\end{tabulary}
\end{table*}

\section{Rankings and experiments}
In this section we elaborate on the results of each track of the relevant editions of SYNTCOMP. All data used for the graphs and analyses given below was fetched from \url{https://syntcomp.react.uni-saarland.de/} for the 2018 edition and \url{https://www.starexec.org/starexec/secure/explore/spaces.jsp?id=329383} for later editions. We have also archived a copy of all scripts and data used in \cite{syntcomp18-21:ds}.

\subsection{Safety track}
Despite the update submitted for the Simple BDD Solver tool, the tool rankings did not change in 2018 compared to 2017. We refer the interested reader to the SYNTCOMP'17 report~\citep{syntcomp17}. In the following years, the safety track has seen neither new participants nor updates to existing tools.

\subsection{Parity track}

\begin{table}
\footnotesize
\centering
\caption{Results of Parity Realizability Track 2020-2021}
\label{tab:parity-real}
\begin{tabulary}{\textwidth}{Lrrr}
      &  \multicolumn{2}{c}{2020} &  \multicolumn{1}{c}{2021}\\
    Tool     & Solved/total & Solving time (s) & Solved/total \\
    \midrule
    Strix & \textbf{122/122}     & \textbf{6.84}             &  150/217       \\
    Knor  & \textbf{122/122}     & 12.71            &      -          \\
    Knor-BDD & (122/122)  & (1.57)          &  \textbf{216/217}         \\
\end{tabulary}
\end{table}

\begin{table}
\footnotesize
\centering
\caption{Results of Parity Synthesis Track  2021}
\label{tab:parity-synt}
\begin{tabulary}{\textwidth}{Lrrrr}
    Tool       & Solved/total & Score \\
    \midrule
    Strix &  260/303 & \textbf{374.89}      \\
    Knor-BDD &  \textbf{276/303}  & 252.66      \\
\end{tabulary}
\end{table}

In 2020 and 2021, two tools participated in the parity tracks: \strix{} and Knor. The former is, oversimplifying, the parity-game solving component of the tool \strix{} that is currently dominating the LTL track. Knor implements several classical and novel parity-game solving algorithms and combinations thereof. In 2020, both tools competed in the realizability track only. In 2021, they competed in the synthesis track only and in a special \emph{combinatorially hard} realizability track. For these two initial editions of the parity tracks, no distinction was made between sequential and parallel subtracks.

\revtwo{2.b}{The results are summarized in {\autoref{tab:parity-real}} and {\autoref{tab:parity-synt}}.} 
Probably the most important insight from the results of these tracks is that 
Knor-BDD --- which only participated \textit{hors concours} in 2020 (since it was submitted after the official deadline), but later became the default configuration of Knor --- outperformed both \strix{} and Knor by almost an order of magnitude only by
switching to a symbolic representation of the input game. Here, participants noted that most benchmarks can be solved easily and that the current bottleneck is parsing the input and constructing an internal representation of the game. This explains the initial advantage Knor-BDD exhibited in 2020. 

An additional point of interest is that Knor-BDD solved \emph{all} benchmarks in 2020; $276$ out of $303$ in 2021; and $216$ combinatorially hard benchmarks out of a total of $217$ in the same year.

\subsection{LTL track}
\begin{table*}
\footnotesize
\centering
\caption{Results of LTL Realizability \& Synthesis Track 2018-2021}
\label{tab:ltl}
\begin{tabulary}{\textwidth}{Lrrrrrrrr}
      &  \multicolumn{2}{c}{2018} & \multicolumn{2}{c}{2019} & \multicolumn{2}{c}{2020} &  \multicolumn{2}{c}{2021}\\
    \midrule
    Tool & \multicolumn{1}{c}{Solved} & \multicolumn{1}{c}{Score} & \multicolumn{1}{c}{Solved} &\multicolumn{1}{c}{Score} &\multicolumn{1}{c}{Solved} &\multicolumn{1}{c}{Score} &\multicolumn{1}{c}{Solved} & \multicolumn{1}{c}{Score} \\
    & \multicolumn{1}{c}{(out of 286)} & & \multicolumn{1}{c}{(out of 434)} & & \multicolumn{1}{c}{(out of 434)} & & \multicolumn{1}{c}{(out of 924)} \\
            
    \midrule
    \strix & \textbf{267} & \textbf{446}  & \textbf{418} & \textbf{717}  & \textbf{424} & \textbf{600} & \textbf{827}  &  \textbf{793}   \\
    BoSy    & 244 & 402 & - & - & - & - & - & - \\
    Party/Kid \& \texttt{sdf}  & 242 & - & - & - & - & - & 730 & 448  \\
    \texttt{ltlsynt}    & 239  & 258 & 361 & 349 & 398 & 360 & 745 & 543  \\
    BoWSer    &  212 & 315 & - & - & - & - & - & -\\
    Otus & - & - & - & - & - & - & 542 & 249\\
    Spore & - & - & - & - & - & - & 499 & -\\
\end{tabulary}
\end{table*}

\begin{figure*}
    \centering
    \includegraphics[width=\textwidth]{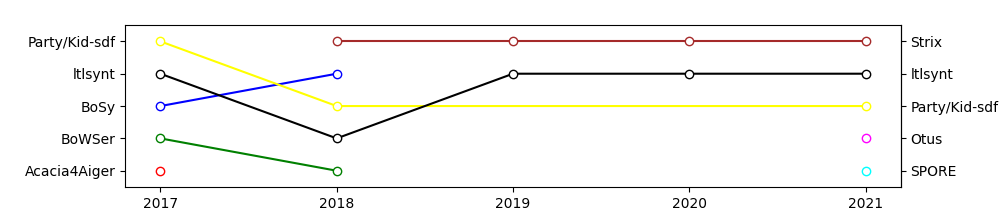}
    \caption{Bump plot of the rankings for the LTL realizability tracks for all editions of SYNTCOMP 2017--2021 (2017 is included for reference since some tools are no longer maintained and being updated)}
    \label{fig:bump}
\end{figure*}

The LTL track has been running since 2016 and the set of \revone{4}{actively participating (i.e. new and updated) tools} has changed a bit since. In \autoref{fig:bump}, we depict the changes in the LTL-realizability rankings every year since 2017, \revtwo{2.c}{and in {\autoref{tab:ltl}} we give an overview of the results in both the realizability and synthesis subtracks from 2018 to 2021}. 

Here, and in similar graphs shown in the rest of this article, we have selected the best configuration submitted per tool. Additionally, all tools --- regardless of whether they were parallel or sequential --- were assigned as score the number of benchmarks solved within the same time limits. It is noteworthy that the best tool in 2017 implemented bounded synthesis algorithms, whereas in recent years the competition has been dominated by parity-game-based tools which focus on optimizing the LTL-to-automaton compilation. Namely, \strix{} has remained first in all rankings since 2018.

\begin{figure}
    \centering
    \includegraphics[width=0.48\textwidth]{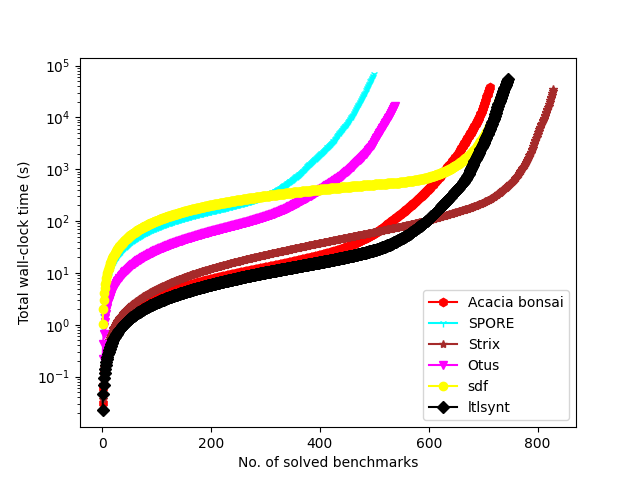}
    \caption{Cactus (a.k.a. survival) plots for the participants of the LTL realizability track of SYNTCOMP 2021; Note that the y-axis is displayed using logarithmic scale}
    \label{fig:cactus-real21}
\end{figure}

Regarding 2021, in \autoref{fig:cactus-real21} we have plotted the total amount of time it takes for each tool to solve increasing numbers of benchmarks. Once more, we do not distinguish between parallel and sequential tools. For reference, we have also included Acacia bonsai in the plot, even if it only participated \textit{hors concours}.

In what follows, we analyze the time and quality rankings of the tools in the synthesis subtrack of the 2018 and 2021 editions of SYNTCOMP. Note that from 2019 to 2020 only ltlsynt and \strix{} participated, and they did so too in 2021, hence our choice of representative years. Additionally, we present the state of the art in terms of scalability for different parameterized families of benchmarks we have used for the competition.

\begin{figure*}
    \centering
    \includegraphics[width=0.48\textwidth]{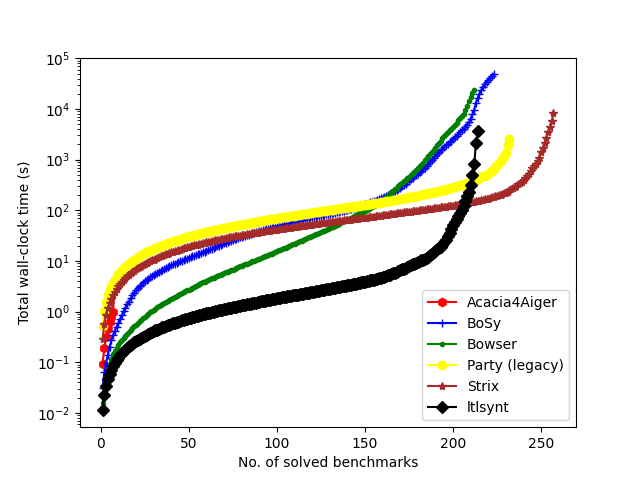}%
    \includegraphics[width=0.48\textwidth]{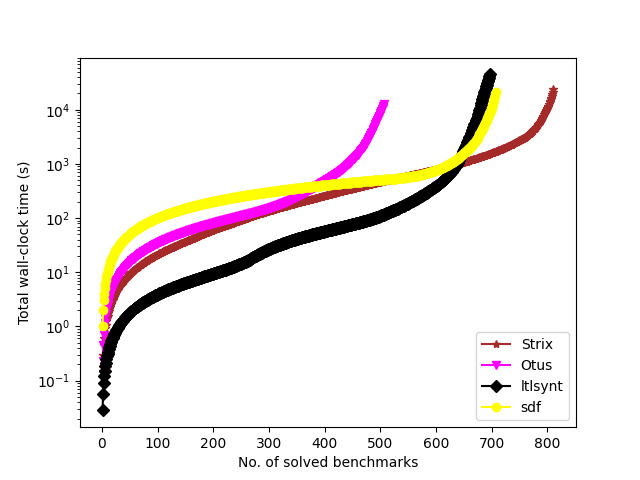}
    \caption{Cactus (a.k.a. survival) plots for the participants of the LTL synthesis track of SYNTCOMP 2018 (left) and 2021 (right); Again, the y-axis is displayed using logarithmic scale}
    \label{fig:cactus-time}
\end{figure*}

\subsubsection{Synthesis subtrack 2018, 2021}
\begin{figure*}
    \centering
    \includegraphics[width=0.5\textwidth]{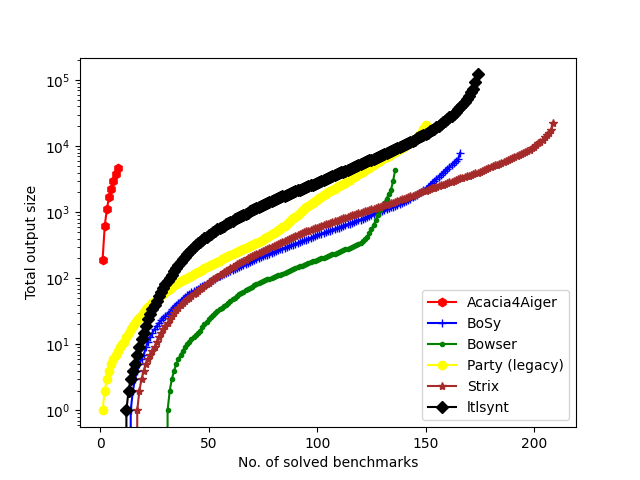}%
    \includegraphics[width=0.5\textwidth]{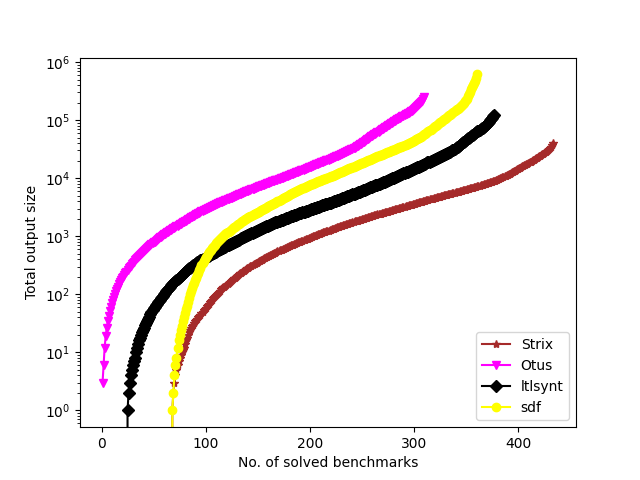}
    \caption{Cactus plots for the participants of the LTL synthesis track of SYNTCOMP 2018 (left) and 2021 (right) --- this time, showing total output size instead of time (counting AND-gates only); Note that the y-axis is displayed using logarithmic scale}
    \label{fig:cactus-size}
\end{figure*}

In \autoref{fig:cactus-time} we have plotted, for each tool, the total amount of time it takes for it to solve increasing numbers of benchmarks. Here, we included configurations from previous years of tools which were updated and Party (as legacy tool) for reference. Additionally, both realizable and unrealizable benchmarks were counted for the score. The previous results should be compared with \autoref{fig:cactus-size}, where we have plotted the total size of the outputs generated for increasing numbers of (realizable) benchmarks.

\subsubsection{Parameterized families of benchmarks}
Presently, we focus on families of LTL specifications defined by TLSF benchmarks with parameters. In 2021, LTL specifications coming from 26 such TLSF files were used in the competition. For most of those families, the results are consistent with the previously summarized results, i.e., the rankings of the tools are preserved when restricting the score to solved benchmarks in that set only. However, some exceptions do exist. 

\begin{figure*}
    \centering
    \includegraphics[width=0.48\textwidth]{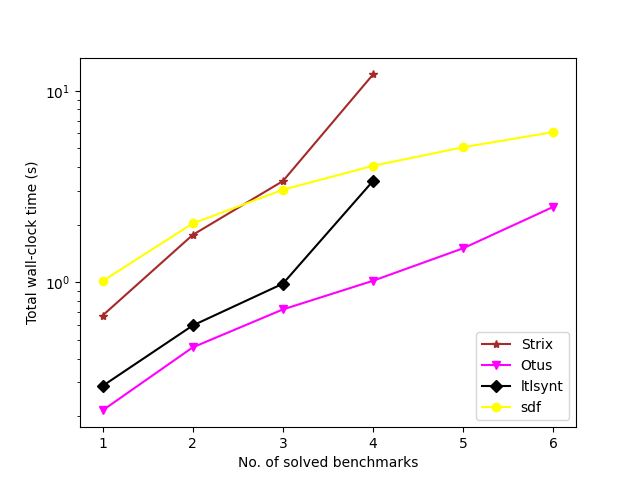}%
    \includegraphics[width=0.48\textwidth]{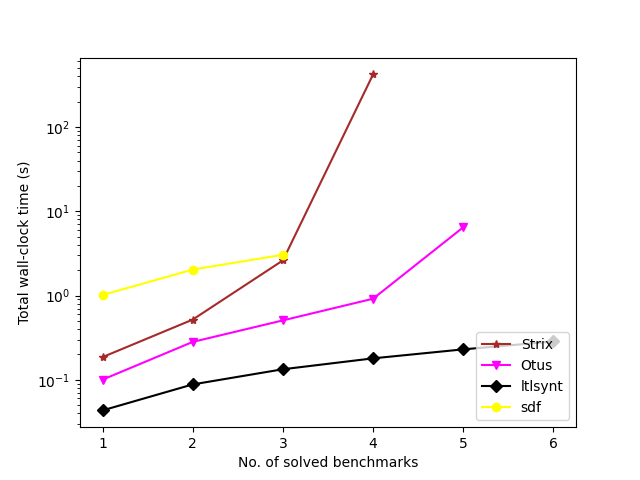}
    \caption{Cactus plots for all the participants of the LTL synthesis track of SYNTCOMP 2021, restricted to the {\tt mux} (left) and {\tt shift} (right) benchmark families}
    \label{fig:comb-logic}
\end{figure*}

\begin{figure*}
    \centering
    \includegraphics[width=0.48\textwidth]{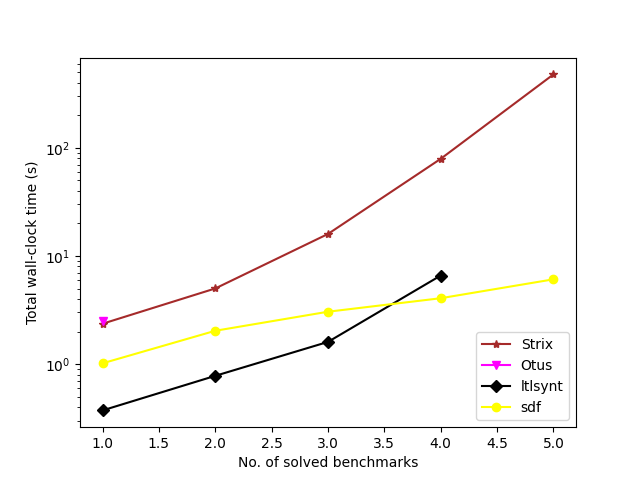}%
    \includegraphics[width=0.48\textwidth]{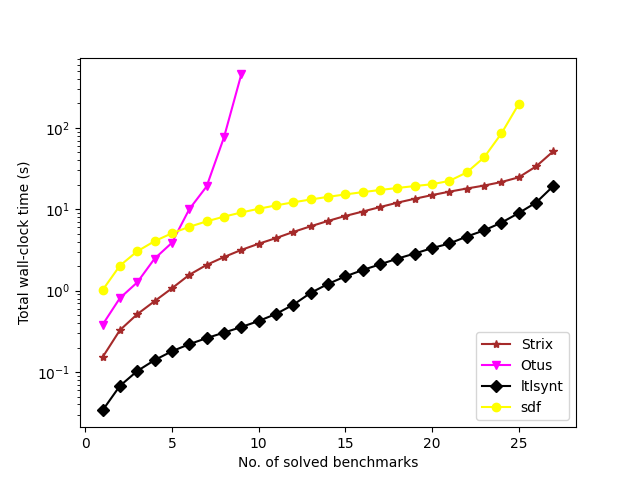}
    \caption{Cactus plots for all the participants of the LTL synthesis track of SYNTCOMP 2021, restricted to the {\tt collector\_v2} (left) and {\tt detector} (right) benchmark families}
    \label{fig:det-coll}
\end{figure*}

\begin{figure*}
    \centering
    \includegraphics[width=0.48\textwidth]{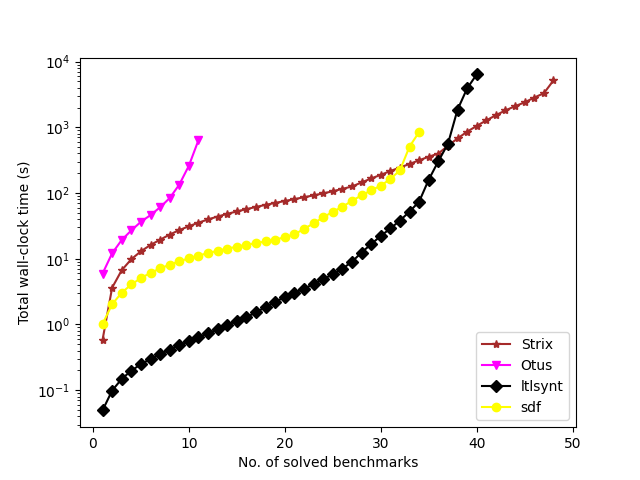}%
    \includegraphics[width=0.48\textwidth]{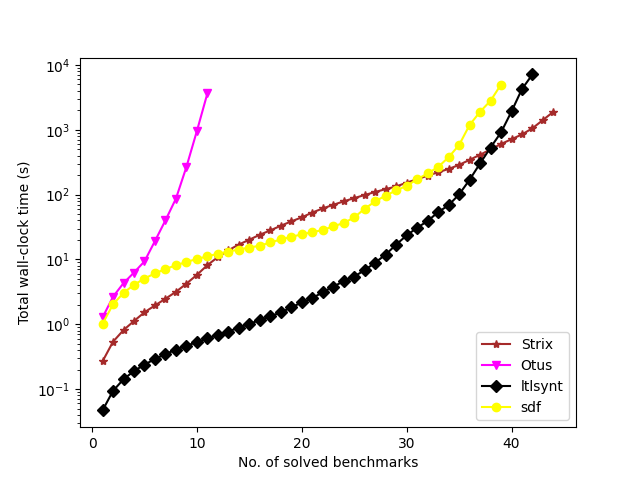}
    \caption{Cactus plots for all the participants of the LTL synthesis track of SYNTCOMP 2021, restricted to the {\tt round\_robin\_arbiter} (left) and {\tt full\_arbiter} (right) benchmark families}
    \label{fig:arbiters}
\end{figure*}

For instance, the families of combinatorial logic specifications {\tt mux} and {\tt shift}, which specify transducers realizing a mutiplexer and a barrel shifter, respectively, are solved most efficiently by tools other than \strix. See \autoref{fig:comb-logic} for the relevant plots.
Other exceptions include the {\tt collector\_v2} and {\tt detector} families of benchmarks, see \autoref{fig:det-coll}. For the former, the  (total) solving time for the best tool is more than order of magnitude smaller than that of \strix. Other interesting families include arbiter specifications such as the {\tt round\_robin\_arbiter} and the {\tt full\_arbiter} families. There, \strix{} is still best overall, but not consistently so, see \autoref{fig:arbiters}.

\subsubsection{Solved benchmark statistics}
Finally, we also give some statistics of the benchmarks solved in the 2021 edition of the competition. We present statistics for the benchmarks solved by all participating tools, those solved by tools implementing the bounded synthesis approach, and those solved by tools implementing the Owl-Pig approach. The data is given in \autoref{tab:solvedltlbenchs}. This should be compared with the data presented in the last row of \autoref{tab:ltlbenchs}.

\begin{table*}
\footnotesize
\centering
\caption{Statistics regarding all LTL benchmarks (not) solved by each solver
class in 2021. The first column indicates the subset of benchmarks under
consideration. Other notational conventions are the same as in
\autoref{tab:ltlbenchs}. The last row is reproduced from
\autoref{tab:ltlbenchs} for comparison.\label{tab:solvedltlbenchs}}
\begin{tabulary}{\textwidth}{Llllll}
   & No. of bench. & Length of LTL formula & Formula depth & No. of inputs & No. of outputs\\
  \midrule
  Solved by $\geq 1$ tool & $869$&$23\leq804, 1538\leq74236$&$0\leq2, 4\leq37$&$1\leq4, 6\leq70$&$1\leq3, 5\leq64$\\
  Solved with b. synth. & $761$&$23\leq720, 1499\leq74236$&$0\leq2, 4\leq22$&$1\leq4, 6\leq70$&$1\leq3, 4\leq64$\\
  Solved with Owl-Pig &
    $842$&$23\leq786, 1276\leq17416$&$0\leq2, 4\leq37$&$1\leq4, 6\leq70$&$1\leq3, 5\leq64$\\
  \midrule
  Not with b. synth. & $181$&$116\leq1403, 4271\leq330512$&$2\leq4, 7\leq37$&$1\leq6, 9\leq65$&$1\leq5, 7\leq64$\\
  Not with Owl-Pig & 
  $100$&$261\leq3282, 8392\leq330512$&$2\leq3, 6\leq17$&$3\leq10, 12\leq37$&$1\leq6, 8\leq30$\\
  \midrule
  All benchs. & $942$&$23\leq896, 2031\leq330512$&$0\leq3, 4\leq37$&$1\leq4, 7\leq70$&$1\leq3, 5\leq64$\\
\end{tabulary}
\end{table*}

\revthree{6}{We observe that the set of benchmarks bounded synthesis tools are able to
solve are longer. Further, benchmarks not solved by bounded synthesis tools have deeper average temporal-operator nesting than those not solved by Owl-Pig solvers. This is in
line with the intuition that complicated LTL formulas may yield complex and
larger deterministic automata, thus slowing down Owl-Pig solvers whose bottleneck
is exactly constructing the automaton. In contrast, recall that
bounded synthesis tools avoid directly constructing a deterministic automaton.
Instead, bounded synthesis tools are mostly affected by how large the bound
$b$ needs to grow in order to find a solution (i.e., how long the satisfaction
of liveness properties has to be postponed, or how many states the smallest
possible solution has).} For instance, the developers of Acacia bonsai report~\citep{abonsai-tacas-23} on the bound required for their tool to solve $667$ of the benchmarks\footnote{To be precise, their experiment also fixes a timeout of around one minute. This means the numbers that follow correspond to a subset of benchmarks that Acacia bonsai can solve very quickly.} from the 2021 edition of SYNTCOMP: From those, $546/667$ finish with $b=2$, and $106$ more with $2 < b \leq 5$, so that only $15$ (of the solved benchmarks) need a bound larger than $5$ and none required more than $b = 8$.

\revthree{6}{Interestingly, based on the statistics of benchmarks not solved by Owl-Pig
solvers, one could conclude that the number of inputs also affects these tools
more than it does bounded synthesis ones.}

\section{Conclusion}
In this article we have reported on the last four editions of SYNTCOMP, the reactive synthesis competition. Furthermore, we analyzed the results of our experimental evaluations, including
a ranking of the participating tools with respect to quantity and quality of solutions. \revtwo{4}{We observe a measurable improvement in terms of the performance of the top solvers in the LTL and parity tracks (see {\autoref{tab:ltl}} and {\autoref{tab:solvedltlbenchs}}, then {\autoref{tab:parity-real}} and {\autoref{tab:parity-synt}}, respectively). In particular, every year we see more tools solving more benchmarks. Additionally, the number of collected benchmarks continues to grow (see {\autoref{tab:ltlbenchs}} and {\autoref{tab:parity-real}}). Also important to note is that the SYNTCOMP benchmarks are now being used in subfields of computer science outside of verification:} In the field of satisfiability testing, synthesis can be seen as an application for \emph{quantified Boolean formula} tools (see, e.g.,~\cite{DBLP:conf/sat/TentrupR19}, where the SYNTCOMP benchmarks are used as a reference). In the field of artificial intelligence, synthesis can be seen as an application for \emph{planning} tools (see, e.g.,~\cite{DBLP:conf/ijcai/CamachoMBM18}). Finally, in the field of programming languages, reactive synthesis tools have been found useful to generate functional programs~\cite{DBLP:conf/haskell/Finkbeiner0PS19}.

\subsection{Lessons learned in LTL synthesis}
In \autoref{sec:class-ltl} we describe a classification of all LTL-synthesis tools into one of two classes. In 2021, \strix{} and Ltlsynt, both in the Owl-Pig class, dominate the LTL tracks. \revthree{7}{We present below a list of ``successful tricks'' which, we believe, have led to the current status of the ranking of the tools.}
\begin{enumerate}
  \item LTL specification rewriting and decomposition (see, e.g.
    \citet{DBLP:conf/lics/SickertE20} and \citep{finkbeiner.21.nfm}, or the
    descriptions of \strix{} and \texttt{ltlsynt} in
    \autoref{sec:updated-tools}).
  \item On-the-fly/incremental or iterative automaton construction and game
  solving on partial automata (i.e. as implemented in \strix{})
\item For bounded synthesis constructing a universal safety automaton:
  \emph{counter reduction}, to reduce the number of states for which counting up to the bound $b$ cannot be reduced to a Boolean flag. 
\item Adaptation of state-of-the-art practical game-solving algorithms (e.g.
  \citep{DBLP:journals/corr/abs-1909-07659,DBLP:journals/corr/abs-2009-10876})
\item Symbolic encodings: (parallel) BDDs, antichains, SAT or QBF
\end{enumerate}
In particular, we observe that the first and second points are (in our
opinion) what allowed \strix{} to take LTL-synthesis to a new level in 2018.
From then onward, they have stayed ahead of other tools by focusing on those
same points together with the fourth one. 
On the other hand, bounded synthesis
with incremental automata constructions or counter reduction based on
rewriting or decomposing the LTL specification has not received as much
attention though it was mentioned as an interesting avenue in earlier
works~\citep{E11unbeast}. 
However, since for bounded synthesis
the construction of the automata is not the bottleneck,
these techniques may not be as valuable.
Additionally, we believe that techniques for finding the right
stratification of the search space by bounding certain paramaters (not only
memory used or visits to rejecting states) offer other avenues of potential improvements,
and have only been explored to a small extent.

Finally, successful tricks seem to be easy to transfer between tools of the same class. As an example, we note that from 2017 to 2018, BoSy integrated the approach that allowed Party/Kid to win in 2017. This allowed BoSy to overtake Party/Kid in 2018. Admittedly, such transfers might be harder between bounded synthesis tools that use different symbolic data structures or encodings.

\subsection{Lessons learned in parity games}
\revthree{6}{From the results of the 2020 and 2021 editions of the competition it is clear that the parity-game tracks require more interesting benchmarks to attract more participants and to make the competition more interesting. Currently, all tools implement similar parity-game solving algorithms and the only advantage Knor seems to have is in terms of the representation of the game and the use of its own (parallel) BDD engine.}

\revthree{6}{For parity games derived from practical
examples, purely symbolic algorithms perform very well; for contrived
artificial constructions, solving an explicit parity game (e.g. in Oink) is superior, though the choice of explicit algorithm makes a big difference.}

\subsection{The future of SYNTCOMP}
In future editions, we will try to replicate the success of the new parity-game track. Indeed, several teams submitting solvers to it eventually ended up extending their tool to participate in the LTL tracks as well. Finally, we will also update some of the rules of the competition so that it remains interesting to the community.

\paragraph{Changes to rules}
Starting with the 2022 edition of the competition, the following changes will be implemented:
\begin{description}
    \item[Benchmark copyright.] All submitted benchmarks will be made publicly available via our repository\footnote{Currently hosted here: \url{https://github.com/SYNTCOMP/benchmarks}} under a CC-BY license.
    \item[Parallel tracks.] To have clearer conclusions regarding which tools are faster, we will no longer differentiate between tool submissions to sequential and parallel tracks. Instead, we will make this distinction by having two rankings: one based on wallclock time (thus favoring parallel configurations), and one based on user-CPU time (favoring sequential ones). Regarding concrete time limits, all tools will be allowed $T$ units of wall-clock time and $4T$ units of user-CPU time (to limit parallel configurations). The factor $4$ was chosen since each \emph{job} ran on StarExec has access to one 4-core CPU.
    \item[StarExec.] For the next 3--5 years, at least, we will continue using StarExec to host the competition. We will follow its evolution in terms of hardware and will later revisit the question of whether alternative (locally hosted) frameworks better fit the needs of the competition and the community.
\end{description}

\bibliography{refs,ltlsyntrefs,sdfrefs,abonsai,sporerefs,strix-otusrefs}

\end{document}